# META-pipe – Pipeline Annotation, Analysis and Visualization of Marine Metagenomic Sequence Data

Espen Mikal Robertsen[1,*], Tim Kahlke[1,#a], Inge Alexander Raknes[1], Edvard Pedersen[2], Erik Kjærner Semb[1,#b], Martin Ernstsen[2,#c], Lars Ailo Bongo[2], Nils Peder Willassen[1]

[1]*Department of Chemistry, Uit The Arctic University of Norway, N-9037 Tromsø, Norway*

[2]*Department of Computer Science, UiT The Arctic University of Norway, N-9037 Tromsø, Norway*

[#a]*CSIRO Health and Biosecurity, North Ryde NSW 2113, Australia*

[#b]*Institute of Marine Research, Nordnes, N-5817 Bergen, Norway*

[#c]*Kongsberg Satellite Services AS, N-9291 Tromsø, Norway*

*\*Corresponding author*

Email: espen.m.robertsen@uit.no

Abstract

**The marine environment is one of the most important sources for microbial biodiversity on the planet. These microbes are drivers for many biogeochemical processes, and their enormous genetic potential is still not fully explored or exploited. Marine metagenomics (DNA shotgun sequencing), not only offers opportunities for studying structure and function of microbial communities, but also identification of novel biocatalysts and bioactive compounds. However, data analysis, management, storage, processing and interpretation are significant challenges in marine metagenomics due to the high diversity in samples and the size of the marine flagship projects.**

**We provide a new pipeline, META-pipe, for marine metagenomics analysis. It offers pre-processing, assembly, taxonomic classification and functional analysis. To reduce the effort to develop and deploy it, we have integrated existing biological analysis frameworks, and compute and storage infrastructure resources. Our current META-pipe web service provides integration with identity provider services, distributed storage, computation on a Supercomputer, Galaxy workflows, and interactive data visualizations. We have evaluated the scalability and performance of the analysis pipeline. Our results demonstrate how to develop and deploy a pipeline on distributed compute and storage resources, and discusses important challenges related to this process.**

## 1 Introduction

In the last decade, the field of metagenomics has exploded due to advances in sequencing technology and data availability. Previously the cost and throughput of the traditional approach, where DNA is cloned into plasmids and sequenced by chain termination sequencing (Sanger sequencing) (1,2), limited the number of samples that could be sequenced. The low cost of high-throughput next-generation sequencing (NGS) allows sequencing more samples to a greater depth. It therefore makes it possible both for small labs and global projects (3–5) to sequence for example environment samples. However, downstream data analysis has become a major limitation for translating the vast amount of sequence data produced to biological knowledge (6).

There are two important challenges for developing metagenomics data analysis pipelines. First, processing and analyzing metagenomics data requires extensive amounts of computation power and storage. Second, sequencing capabilities increase exponentially, at a much higher rate than advances in computation power and storage capacity. Closing this gap requires faster algorithms implemented to run efficiently on large scale-computer systems (7).

Previous analysis resources have demonstrated their usability for metagenomics data analysis, including EBI-Metagenomics (8,9), Metagenomics-Rapid Annotations using Subsystems technology (MG-RAST) (10) and Integrated Microbial Genomes and Metagenomes (IMG/M) (11). However, these are not developed for the marine metagenomics domain and do not offer the extensive annotation options, flexibility and visualization needed to pick interesting targets for further investigation. In particular, there is a need to produce full-length annotated genes from metagenomic assemblies. In addition, these are typically run on a server administered by a single organization, often resulting in scalability problems for free to use resources, or costly fees for pay to use services. There is therefore a need to develop a scalable pipeline for the marine metagenomics field. To ensure fast development, scalability to flagship projects, and easy deployment of the pipeline, it should utilize existing frameworks and infrastructure resources and services when possible.

In this paper we describe META-pipe, an automated pipeline for annotation and analysis of metagenomic and genomic sequence data. We provide META-pipe as a service targeted for marine metagenomics (Section 2.1). We have integrated META-pipe with Galaxy (12–14), our university supercomputer, and we use national storage and authentication resources (https://nels.bioinfo.no/) (Section 2.2). We briefly describe an initial evaluation of the quality of the analysis results (Section 3.1), and we have evaluated the scalability of META-pipe and the overhead of the infrastructure integration (Section 3.2).

We make the following contributions:

- We describe the complete META-pipe pipeline and its functionality.
- We describe our approach for integrating the customized META-pipe workflow manager with Galaxy and the job management systems on a Supercomputer.
- We demonstrate how our pipeline uses infrastructure services for data storage and processing, and authentication and authorization services.
- We evaluate the performance and scalability of META-pipe and discuss limitations that will be addressed in future versions of META-pipe.

Summarized our findings show how to develop, scale, and deploy a domain specific pipeline on local and national infrastructure resources. We also identify several issues with the current pipeline and its implementation and outline directions for future work.

## 2 Methods

We describe the choice of tools in META-pipe and the implementation and infrastructure integration of the pipeline.

### 2.1 Analysis pipeline design

META-pipe efficiently produces full-length annotated genes from metagenomic assemblies, and offer the extensive annotation options, flexibility and visualization needed to pick interesting targets for further investigation. Initially, we developed a coding sequence functional analysis pipeline (coined "GePan" at the time (15)) and integrated it with META-rep for effective visualization of results. Subsequently, we added modules for pre-processing and taxonomic classification to provide a more streamlined and complete analysis.

At present, META-pipe provides taxonomic and functional analysis for both 16S amplicon data and whole genome shotgun sequence data. It supports high throughput sequencing data and provides assembly, focusing the analysis on full-length genes. The pipeline consists of three major modules, pre-processing, taxonomic classification and functional analysis (Fig 1). All modules are available as individual workflows, except for assembly in pre-processing, which is run manually on either a large memory computer or our a supercomputer. Workflows can be tailored to the specific needs for the analysis of a sample and it is also possible to add additional steps or to omit some of the steps.

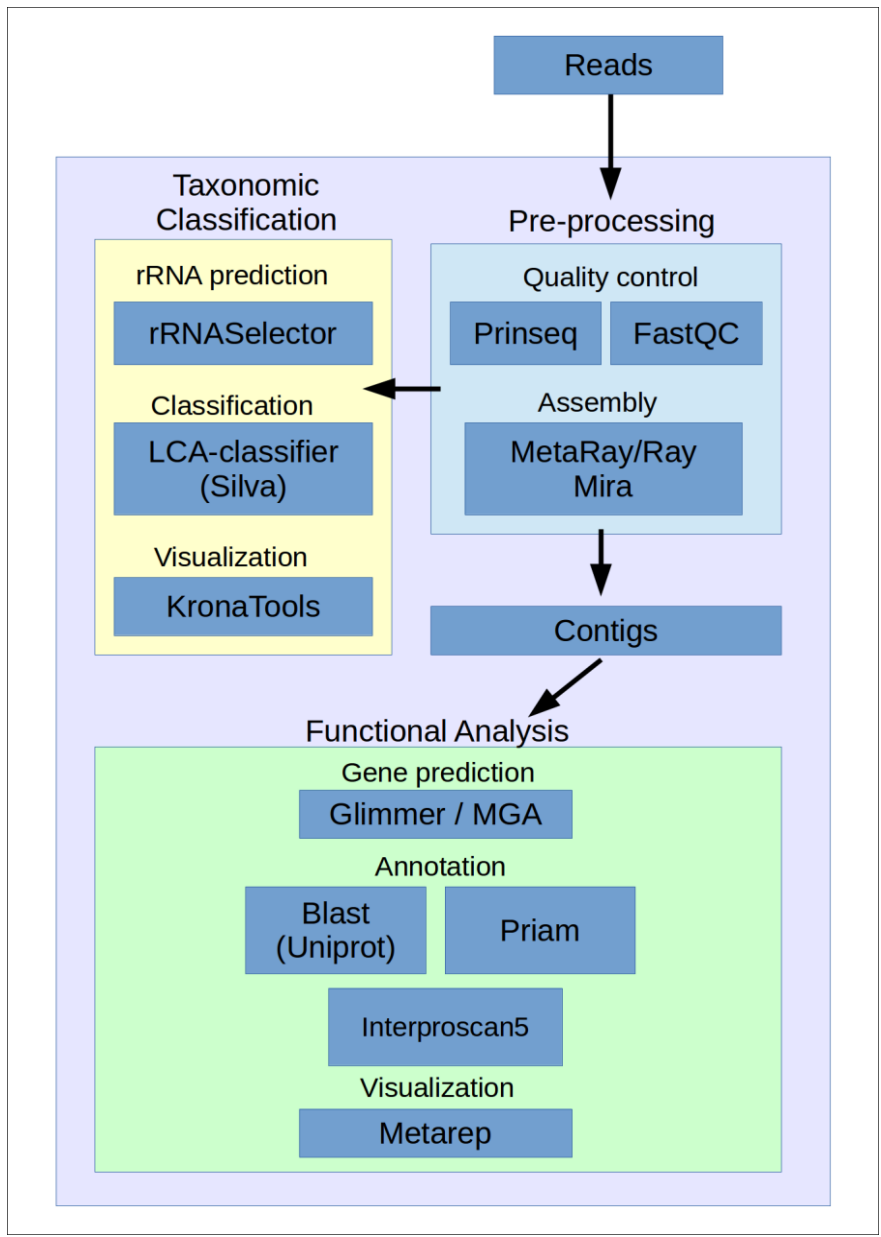

**Fig 1: A schematic overview of the components and major modules of META-pipe.**

### 2.1.1 Input data and quality control

META-pipe accepts raw sequencing data as initial input. Before analysis, META-pipe pre-processes raw reads to remove dubious and low quality data using Prinseq (16), either with manually chosen parameters or with parameters that are specifically tailored for the utilized sequencing platform, such as Illumina. The user can inspect the results of the filtering process using graphs from FastQC (17).

### 2.1.2 Taxonomic classification

The taxonomic classification module uses the filtered sequences to produce an overview of the organisms present in a dataset. Initially, the module uses rRNA selector (18) to predict and trim rRNA sequences from reads. Predicted sequences are blasted against a curated version of the SILVA database (19) (shipped with LCA-classifier) which are then fed to the classification program LCA-classifier (20). It uses the "least common ancestor" algorithm to classify reads into operational taxonomic units which are displayed using the interactive pie charts produced by Krona (21).

META-pipe users can also use the taxonomic classification module to mask predicted 16S rRNA sequences from the dataset, excluding them from the assembly stage and thus reducing spurious contigs (contigs are longer segments of DNA from the sample) from assembly. The module can also classify 16S amplicon datasets, but quality checking and filtering must be done manually beforehand as this is not a part of the workflow for this particular dataset type.

### 2.1.3 Assembly

META-pipe assembles the dataset to produce a set of contigs for functional analysis. Assembly of metagenomic datasets typically requires substantial amounts of memory due to the complexity and sequencing depth required. The memory usage can be hundreds of gigabytes, depending on the assembly program used and the size and type of the input dataset. To address this challenge, META-pipe offers two different assembly programs. Mira (22,23) for genomes and smaller sequence datasets running on a single node, and Ray/MetaRay (24,25) for larger metagenomic sequence datasets where computation and in-memory data structures are distributed among multiple nodes, and inter-node communication is implemented using the Message Passing Interface (MPI) (26)

### 2.1.4 Functional analysis

The input to the functional analysis module is the contigs produced in the assembly step. This module is based on the GePan annotation pipeline for genomic sequence data. We extended the functionality and scalability of GePan to support metagenomic dataset analysis. First, the META-pipe functional analysis module predicts a set of genes from the contig input sequences provided, using Glimmer (27,28) or MetaGeneAnnotator (MGA) (29,30). This set is split relative to the number of CPU-cores utilized, in order to effectively scale with the computational resources available. Secondly, it runs selected available tools on each dataset split in parallel. META-pipe offers the tools Blast (31) against the UniprotKB database (32), Priam (33) to identify enzymes and Interpro (34,35), which offers an additional eleven signature databases queried using Hmmer (36). Each tool produces a set of output files that are parsed by an annotation module, producing an intermediate file conjoining predicted genes with their respective functional annotations.

### 2.1.5 Visualization of analysis results

The annotation module exports the produced output data for visualization using standalone programs. META-pipe provides XML, EMBL, and a tab separated text format, as well as output files in the custom format used by the Metarep visualization tool (37). Metarep is a web-interface open source tool for high-performance comparative metagenomics visualization. It allows for viewing, browsing and comparing datasets with built in tools such as statistical tests, multidimensional scaling heatmaps and pathway analysis. Additionally, our modified version of this visualization tool supports sequence retrieval making it possible to download and further investigate particular genes of interest.

## 2.2 Implementation, integration, and deployment

We have designed META-pipe to scale to large-scale metagenomics datasets. We have integrated META-pipe with Galaxy and Metarep to provide a flexible but well known user interface. In the Norwegian e-Infrastructure for Life Sciences (NeLS) project (38), we have deployed META-pipe on national infrastructure resources to make it easy for user to access META-pipe and to take advantage of the national storage (NorStore) (39) and high performance computing resources (Notur) (40).

In this section we describe how we integrated META-pipe with Galaxy, Metarep, our local Notur Supercomputer (Stallo) (41), and how we use the NorStore infrastructure resources for storage, and the FEIDE services for authentication and authorization (42). We believe many of the challenges and solutions apply to pipelines implementing using other workflow managers, visualization tools, and infrastructures. We are currently integrating META-pipe with European infrastructure resources in the ELIXIR Excelerate project (43).

### 2.2.1 Use of META-pipe

META-pipe is one of several pipelines in the Norwegian e-infrastructure for life sciences (NeLS). NeLS services use a standardized user interface for login, data submission, data management, and workflow management. Below we describe how a NeLS user uses META-pipe (Fig 2).

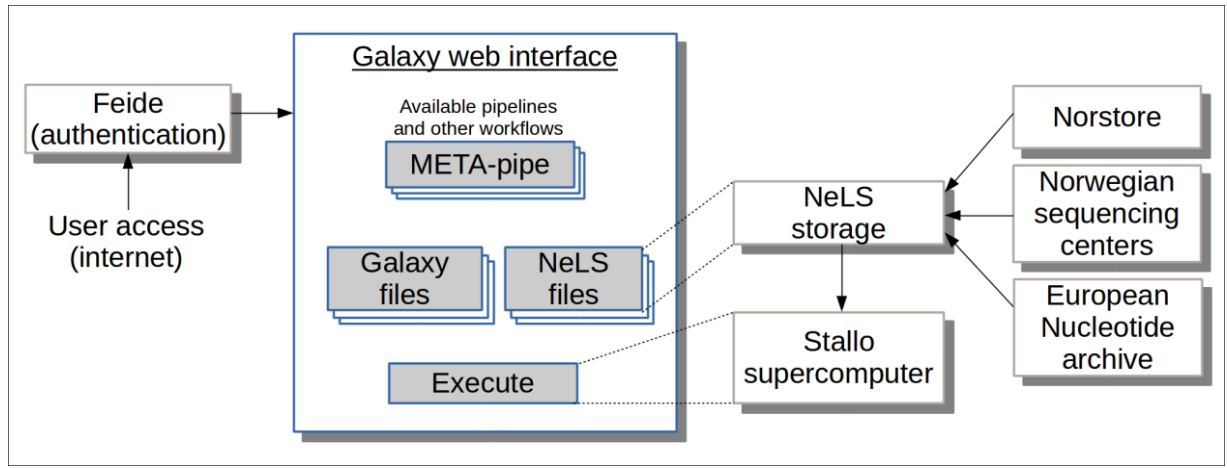

**Fig 2: META-pipe integrated with the Norwegian eInfrastructure for Life Sciences (NeLS) services.**

First, the user opens the NeLS webpage (https://galaxy-uit.bioinfo.no/), and authenticates by providing their Feide (https://www.feide.no/) username and password. Feide is a federated identity provider for all employees and students at Norwegian universities, hence all Norwegian academic users can use their default login username and password. Other users can request access through a separate NeLS identity provider (NeLS IdP).

After login, the user is redirected to our NeLS Galaxy web frontend that provides three options for uploading the data to be analyzed. First, the data can be uploaded from a local file on the user's computer through the browser. Second, the data can be directly downloaded from a provided URL. Third, for large files, the data can be copied to the user's project area in the NeLS storage system, which is connected to the same high-bandwidth network as the storage systems used by the Norwegian sequencing labs and supercomputer centers. The project area files are visible from Galaxy, and the files can be added to Galaxy workflows. The files uploaded through the browser can also be stored in the user's project area.

To analyze the data, the user selects the META-pipe tool in Galaxy and then configures the pipeline parameters such as which tools to run, input files, reference database versions, and output formats. Once the workflow is configured, the user presses the execute button in Galaxy to execute the pipeline in the background. This will create a history pane in Galaxy where the user can view the current status of the job. Currently queued or running jobs are colored yellow, and completed jobs are colored green. When the job is done, the user can examine the output data in the galaxy view panel and download the files to their own computer.

### 2.2.2 Galaxy workflow

We use Galaxy as the user interface for configuring and executing META-pipe. Galaxy is an open sourced web-based platform where users can create and share workflows for easy usage of bioinformatics tools and applications. It provides a graphical user interface that does not require using programming or command line tools. Galaxy was chosen as the common user interface in NeLS, since we believe it is the most popular, and hence most familiar, workflow manager for bioinformatics. In addition, some of the partners in the NeLS project already had previous experience using Galaxy for the Lifeportal (44).

The META-pipe workflow consists of the pre-processing tools, taxonomic classification tools and a single Galaxy tool for the functional analysis. This tool outputs results into a single zip file that contains all functional annotation files depending on the output format selected. Taxonomic classifications can be viewed with Krona-charts in the browser, or downloaded for use with standalone programs.

We implemented the functional analysis as a single Galaxy tool since we already had a script based workflow manager for parallel cluster execution. The single Galaxy tool approach has the disadvantage of hiding the individual functional analysis tools from Galaxy and hence the user. However, a Galaxy workflow with all functional analysis tools visible requires a redesign of the cluster job execution of the functional analysis pipeline.

We run our Galaxy instance on a virtual machine with the following specifications: 32GB RAM, 2 virtual cores and 4TB disk space. We can have a lightweight virtual machine since we offload compute intensive tasks to the Stallo super computer as described below. However, we need storage for the user's data as described in the next section.

### 2.2.3 Storage infrastructure integration

We have integrated our Galaxy VM with the storage systems on our Stallo Supercomputer and the NeLS storage system located at another site in Norway (Bergen) (Fig 3). All three storage systems provide project storage during the project. For long-term (archival) storage, Norwegian users should use the services provided by the NorStore storage system, so we do not address the many challenges and solutions for long-term data scientific data management.

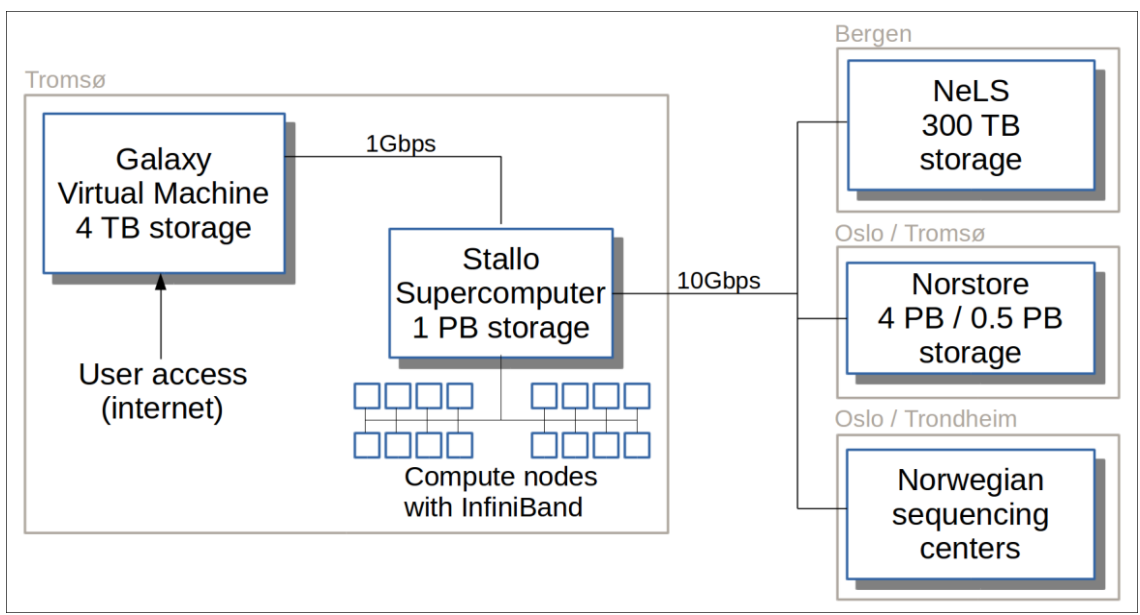

**Fig 3: Storage architecture.**

The dataflow for users that wish to archive their input and analysis results is as follows:

1. Deposit the raw data in the NorStore archive. The data may be copied directly from temporary storage at a Norwegian sequencing center over a high bandwidth network.
2. Copy the raw data to the user's projects space in the NeLS storage system.
3. Copy the data from NeLS storage into Galaxy using Galaxy tools developed in the NeLS project.

Then the Galaxy workflow will:

4. Copy the data from Galaxy to the Supercomputer storage system, from where the data is read by the compute nodes.
5. Execute the Meta-pipe workflow on the Supercomputer.
6. Copy the results back to Galaxy.

Finally, to save the results, the user must:

7. Copy the results to NeLS storage.
8. Archive the results in NorStore.

Users that do not wish to archive their raw data or analysis results may skip steps 1-3 and 7-8, and instead copy the input data directly to Galaxy, and then download the results when the analysis is done.

We treat the data in the users Galaxy directory as the master copy, and the other copies as replicas. It is therefore up to the user to ensure that the data is archived for long-term storage. Galaxy functions as the main data repository and as data traffic needs to pass through Galaxy, this approach has a performance overhead since the data must be copied to and from a virtual machine with limited bandwidth. A more efficient solution is therefore to copy the data using Secure Copy (SCP) directly from NeLS storage to the Stallo storage system over a high-bandwidth network link. The NeLS storage system already supports this, and we plan to use it in our next version of META-pipe.

### 2.2.4 Supercomputer integration

We have integrated the META-pipe scripts with Galaxy and the Torque job management system on the Stallo supercomputer (Fig 4). The virtual machine we run Galaxy on does not have the compute resources for all stages in META-pipe. Stallo has 824 nodes, 18.144 cores, 26.2TB DRAM, 2 PB centralized storage, InfiniBand 40 Gb/s (4x FDR10) interconnect, and a software stack consisting of Rocks 6.0 (CentOS 6.2), the Lustre file system, and the PBS/Torque queuing system. Stallo provides the resources needed to analyze the largest marine metagenomics datasets, and to execute several analyses in parallel. Additionally, we also have a local workstation with 12 cores, 256GB DRAM and 5.5TB storage for metagenomic assembly of small datasets.

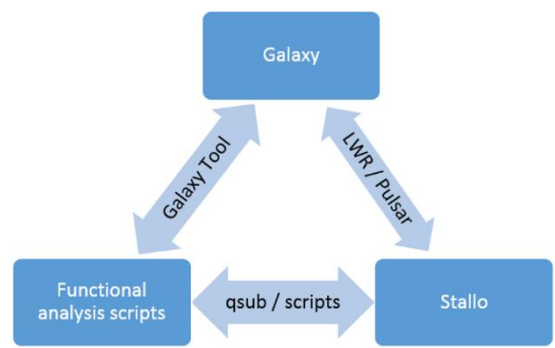

**Fig 4 Supercomputer (Stallo) integration architecture.**

A Supercomputer (high performance computing cluster) such as Stallo typically consist of many compute nodes, connected with a fast interconnect, and a centralized storage solution. The platform provides a network file system for storage, and a parallel job execution system such as Open Grid Engine (OGE) (45) or Torque (46). For parallel execution of META-pipe on Stallo, we must manage data in the global file system, partition the data processing, and submit jobs to the job management system.

For the functional analysis tools in META-pipe we implemented a workflow manager that handles data management and parallel job execution. As is common, we implemented our workflow manager using a set of scripts. As is also common, the size of the scripts grew, resulting in our current implementation that has about 10K lines of Perl code. We note that implementing a pipeline manager from scratch requires solving several issues, including failure handling, provenance management, user interface, and portability. We are addressing these issues in ongoing work.

We use the LWR (now renamed to Pulsar) Galaxy services to execute META-pipe tools in parallel on the Stallo Supercomputer. LWR is used to submit compute intensive Galaxy tools to Stallo. LWR communicates with Galaxy via the Galaxy API and a RabbitMQ (AMQP) message queue (Fig 5). We are currently running two instances of LWR. The first is configured to run Meta-pipe. The second is configured to submit compute intensive Galaxy tools to Stallo over DRMAA (Distributed Resource Management Application API). The large memory workstation is currently not accessible via Galaxy/LWR. We can make it accessible by installing LWR on it and add it to the Galaxy configuration.

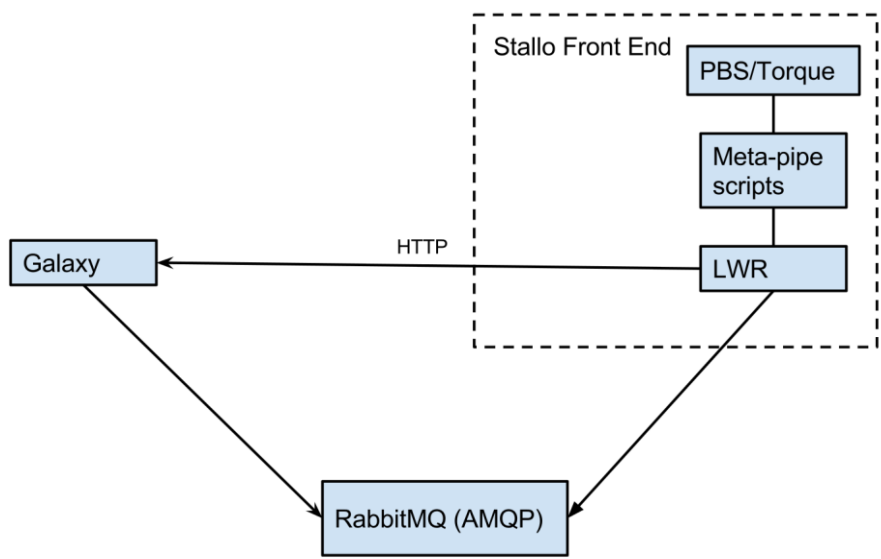

**Fig 5: META-pipe/Galaxy/LWR setup**

To execute the functional analysis, the user clicks "submit" in the Galaxy user interface. Galaxy will submit the job to LWR using the message queue, which in turn executes the specific galaxy tool for META-pipe functional analysis. All required parameters (the reference database versions to use, tools to run, input files, and desired output format) are specified in this tool. When executed, this tool runs a script that handles error checking during a job, which in turn executes the wrapper script for META-pipe functional analysis. The wrapper creates temporary directories, submits tool jobs to PBS/Torque with selected parameters, saves results and deletes temporary files. Once the jobs are complete, LWR transfers the data back to Galaxy for the user to inspect.

### 2.2.5 Metarep integration

Currently the data exploration and visualization tools for metagenomics in Galaxy do not offer the functionality required for META-pipe. We have therefore integrated META-pipe with the Metarep metagenomics visualization tool. To use Metarep, the user configures the Galaxy workflow to output results in the defined import template used by Metarep. Once result files are downloaded from Galaxy, these files can be imported using the provided SolR import scripts shipped with Metarep.

We also modified Metarep (version 1.4.0) to add features that allow the user to click on a specific gene's ID in the view and search panels to display a window with the genes nucleotide sequence as well as it parent sequence (full length contig or genome) in fasta format. We also added a batch download function to download a set of genes in multi fasta format that matches for example a keyword search against a specific annotation (all available at https://github.com/emrobe/META-pipe). With this added functionality, users can in addition to browse and view their results, and also download sub sets of genes for further analysis or exploration with other biological applications. We store the sequence data as formatted blast databases in a directory hierarchy that matches the project ID and sample ID layout used by Metarep. Sequences are retrieved from their respective directories and displayed/downloaded in the browser using the blast program *Fastacmd* on demand.

## 3 Results

In this section we focus on the scalability and performance of META-pipe. We provide answer to the following three questions:

1. Can META-pipe efficiently utilize the resources on a Supercomputer?
2. Does META-pipe scale with respect to data set size and resource usage?
3. What is the overhead and bottlenecks of our infrastructure integration?

The quality if the META-pipe analysis results are evaluated in (under preparation for publication). We summarize the results below as a use case for META-pipe.

### 3.1 Biological use case

In a pilot project within ELIXIR (under preparation for publication) we compared the analysis results of META-pipe with the results from the EBI metagenomics portal (pipeline version 2.0). We could do the comparisons since we have interoperability in functionality and data formats for these two metagenomics pipelines. We used four metagenomic datasets. Two of the datasets were marine samples taken from the Barents Sea, while the other two were intestine samples from moose and sea urchin from locations in northern Norway. We found that META-pipe results had a richer taxonomic classification and a functional analysis better suited for analysis of full-length genes. This is especially beneficial for projects where a user wants to produce data for further exploration in biological applications, for example when mining for novel enzymes to be included in new commercial products. A YouTube video summarizing this project is available at https://youtu.be/uSsvIZhY8Hs

### 3.2 Pipeline scalability and performance

In this section we evaluate the scalability and efficiency of the functional analysis module of META-pipe and the overhead of the infrastructure resource integration. We do not evaluate the scalability of the taxonomic classification module since it has much smaller execution time since only does a single blast step with a small database (Silva), and since blast-scalability is evaluated for the functional analysis module. Our goal is that META-pipe should scale to the dataset sizes of marine flagship projects such as TaraOcean (5) and Ocean Sampling Day (4), and that several pipeline instances can be run at the same time for smaller projects. There are four parts that influence the scalability and performance of META-pipe:

1. The scalability of the individual tools in the pipeline. If a tool does not scale, the pipeline will also not scale. If a tool does not scale, we must choose a more scalable tool. If the tools scale well, we can reduce analysis time by using more resources.

2. The scalability and resource utilization of the parallel META-pipe implementation. The pipeline implementation should take care of data dependencies in the pipeline, and efficiently schedule the tools for execution.
3. The overhead of the Galaxy integration and the distributed storage system. The overheads should be low compared to the overall pipeline execution time.
4. The policies of the Supercomputer. The pipeline execution time is influenced by for example the queue wait time and resource allocation policies. We cannot change these for our experiments, but we can measure their impact on the execution time and we can easily predict the effect of policy changes to the analysis completion time.

For the evaluation we run META-pipe on the Stallo Supercomputer. We used 32, 64 or 128 of the 18.144 cores. Our jobs allocated one core per process (32, 64, or 128). We cannot choose the nodes allocated for a job or the number of cores allocated per node, but we assumed that the process to node mapping does not influence the execution time of our jobs. This assumption does not always hold as discussed below. When a job is submitted to the queue, it is blocked until the queue system allocates the requested resources. This is done in a semi-round-robin system, with additional priority given to smaller jobs, and for users that have few jobs running.

For the evaluation we use the "Muddy" (European Nucleotide Archive sample accession: SAMEA3168559) marine metagenomic sediment sample from the Barents Sea. This dataset is representative for medium sized (9 GB) and high complexity marine metagenomics dataset that we expect most of META-pipe users will analyze. Note that the performance and hence scalability of the tools may depend on the input data complexity, size, quality, and the sequencing technology used to generate sequence data. Complex datasets which are rich in terms of unique organisms present and abundance, and that have large size and low quality base calling, will increase memory usage and affect performance in de Bruijn graph assemblers such as MetaRay. Different sequencing technologies will also contribute with their respective sequencing traits, such as average read length and unique sequence quality flaws.

We evaluated the scalability of functional analysis tools with respect to dataset size by choosing a different cutoff length for the results from MetaRay. The "Medium" dataset, which we have used for the scalability experiments, has a cutoff of 300 nucleotides, and a size of 21 MB. "Small" has a cutoff of 400 nucleotides, resulting in an input size of 12 MB. "Large" has a cutoff size of 250 nucleotides, resulting in an 34 MB input file.

For the MetaRay experiments the input data was stored on the shared file system on Stallo. For the Meta-pipe experiments the input data was loaded from Galaxy.

### 3.2.1 Tool scalability

The first step in META-pipe with a significant execution time is assembly. We evaluated the scalability of MetaRay on Stallo using the "Muddy" sample with an average from 9 assemblages, 3 using 32 cores, 3 using 64 cores and 3 using 128 cores (Fig 6). MetaRay does not scale well on Stallo, and it becomes an increasingly larger contributor to the total execution time with more cores. We have not been able to tune the MetaRay parameters to improve scalability.

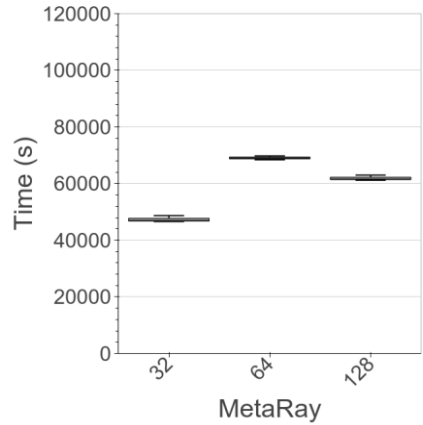

**Fig 6: MetaRay scalability on Stallo using 32, 64 and 128 cores for the "Muddy" sample.**

We have also evaluated MetaRay scalability on our *ice* development cluster (8 nodes with 4 cores and 32 GB DRAM) and on Amazon EC2 (20 virtual nodes with 8 cores and 32 GB DRAM). MetaRay had similar performance issues on these platforms. More surprising, our results show that (the MPI based) MetaRay does not benefits from the lower network latency and higher bandwidth of the Infiniband interconnect on Stallo compared to Ethernet on *ice* and the virtual cluster on EC2. Our initial results also show that no other tool significantly benefits from the networking infrastructure present on the Stallo HPC architecture.

The execution time breakdown for the functional analysis META-pipe tools is shown in **Error! Reference source not found.**. Each bar in this graph is based on the median percent of runtime (tasks) from the median of 3 tool runs (jobs). With a small number of cores (32), and for the small dataset, META-pipe execution time is dominated by Interproscan and Blastp, with 80 % of the total run time combined. However, these scale well (Fig 8) both in terms of cores used and increasing dataset sizes, and hence with more cores these contribute less to the overall execution time.

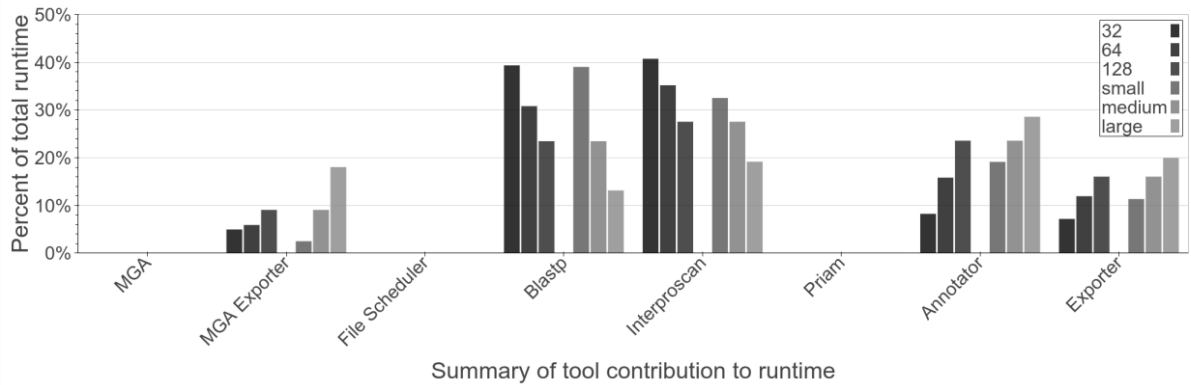

**Fig 7: Execution time breakdown for the functional analysis tools.**

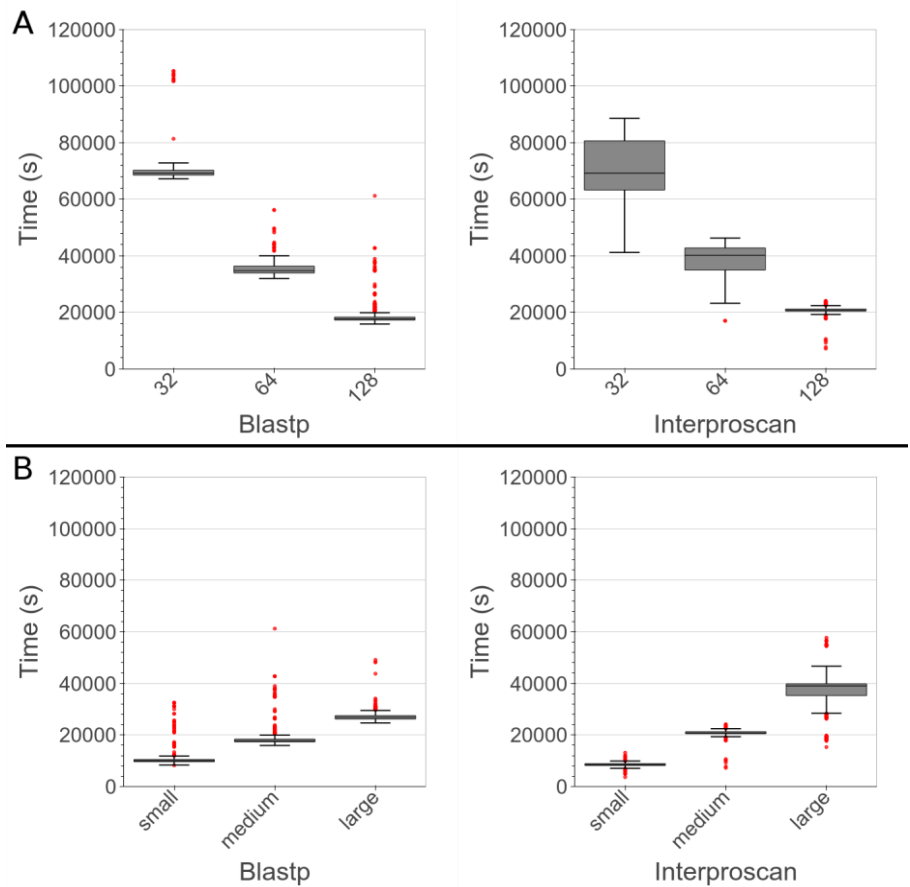

**Fig 8: Scalability evaluation of Blastp and Interproscan with regards to number of cores (A) and dataset sizes (B). Outliers are marked in red.**

Our annotator and two exporter scripts does not scale well, and they contribute to 50% of the functional analysis execution time using 128 cores. In addition, with increasing dataset sizes the computation time exponentially increases (Fig 9). However, this is due to an inefficient implementation. We re-implemented these tools (but did not integrate these with the other tools in the pipeline), and reduced the execution times on the medium dataset with 128 cores to respectively 1 second for the MGA exporter, and 5 seconds for the Annotator and Exporter that we combined into a single tool running on a single node (the re-implemented tools are available at https://github.com/emrobe/META-pipe).

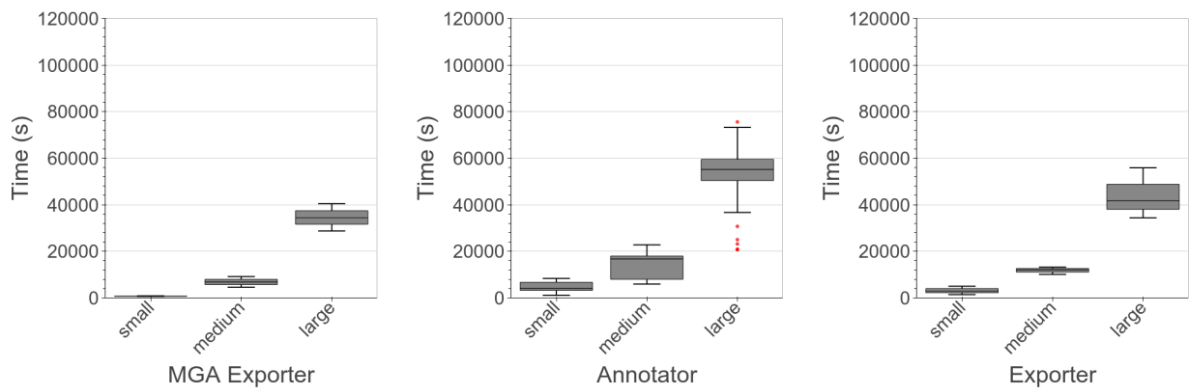

**Fig 9: Execution time with increasing dataset sizes for the tools that do not scale well. We have re-implemented these to improve scalability.**

We used the median task execution time to visualize execution time breakdown since we experience occasional problems caused by straggler nodes in some runs on Stallo (Fig 10). On some nodes, specific tasks may run up to 3-fold longer than their siblings. The increase on task execution time on these nodes increases the job execution time since it is determined by the slowest task. We only experienced such perturbation on Stallo and we have verified that the tools have balanced workloads We could not control the number of cores used per node, or which nodes to use, since the process to node mapping was determined by the Torque queuing system on Stallo.

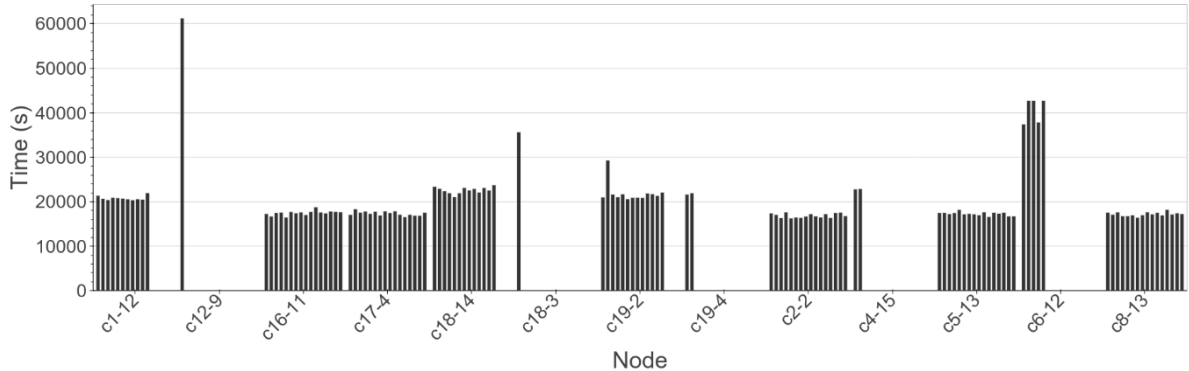

**Fig 10: Task execution time for BLASTp on different nodes for one of the 128-core medium-size runs. The work per task in BLASTp is load balanced so the differences in execution time are caused by perturbations on the node, which results in overall longer execution time for the parallel job.**

In conclusion, the META-pipe tools that contribute most to the execution time scale well. The one exception is MetaRay, that did not scale in our evaluation, and that requires a significant effort to optimize. In addition, straggler tasks increase parallel tool execution time, but we can reduce their impact by increasing the number of tasks (by decreasing the size per task). This will result in fewer task mapped to a straggler node and lower execution time for straggler tasks.

### 3.2.2 Pipeline scalability

A significant contributor to the overall META-pipe completion time is the time the META-pipe jobs wait in the cluster scheduler queue. The wait time has a very large variation depending on the load on Stallo. We have observed wait times ranging from less than a minute up to several days. Large wait times can be expected on a Supercomputer with a maximum resource utilization policy. A large variation in job completion also makes it hard to give a good estimate for the job completion time for users. A policy with shorter and more predictable wait time would require a platform with spare capacity to immediately allocate the needed resources. Such a policy can be provided by either a dedicated cluster, dedicated part of a Supercomputer, or by cloud IaaS (Infrastructure as a service) providers such as Amazon and Google.

Although a META-pipe analysis is submitted as several jobs run in sequence, the cluster Scheduler is configured such that these jobs are mostly executed without any wait time between the jobs by ensuring that the next job is run on the same allocated resources.

Our results do not show any performance reduction when six instances of META-pipe are run on different portions of Stallo. This is expected, since we only use a small fraction of Stallo for each instance. Hence, we can easily support several concurrent users provided that the needed resources are allocated at the same time.

In conclusion, our parallel implementation of META-pipe scales well and the pipeline management has no significant overheads. It also supports many concurrent META-pipe analysis instance. But the job completion time is limited by the queue wait time policy on Stallo. Our ongoing work addresses the queue wait time limitation.

### 3.2.3   Integration overhead

We measured the overhead of the Galaxy integration by measuring the time spent from starting the META-pipe workflow in Galaxy to the job is submitted to the queue. We used a small input file of 30 MB and a larger 1 GB input file. These are representative for META-pipe datasets respectively after and before assembly. For the small file the overhead is a few seconds. For the larger file, the data copy overhead is about 90 seconds. For both input file sizes the Galaxy integration overhead is insignificant compared to the total pipeline execution time.

We simulated up to 12 concurrent Galaxy users by submitting many jobs at the same time without any noticeable effect on the responsiveness of Galaxy or job runtime.

Our results show that the integration overhead is non-significant when compared to the analysis time, and that multiple concurrent users can run jobs on META-pipe.

## 4   Discussion

### 4.1   Choice of pipeline tools

We had two criteria for choosing the META-pipe tools: the quality of the results and tool scalability. We are continuously evaluating new tools and databases either as replacement or additions to META-pipe in order to improve the quality of the analysis results.

Our results demonstrate that the META-pipe tools scale well, with exception of MetaRay. We have examined several alternatives to MetaRay. Similar to the results reported in Spaler (47), we did not find a parallel assembler that scales well. We believe the scalability issue in MetaRay is due to its implementation of parallel graph processing. This is a recognized problem in large scale graph processing and several specialized graph processing systems, such as Pregel (48) and the Graph X (49) library for Spark have been developed to solve these issues. The latter is especially interesting since our new version of META-pipe will be implemented in Spark. The Spaler assembler is implemented using Graph X, but it is not currently publically available, so we have not been able to test its scalability and the quality of the results for marine metagenomics datasets.

### 4.2   Parallel pipeline execution

The script based parallel implementation of the functional analysis in META-pipe scales well and has good performance. However, it has several issues for a production system that we are addressing in ongoing work.

First, the scripts were implemented for the OpenGrid job scheduler and the implementation made many assumptions about OpenGrid such as environment variables, the syntax of the qsub command, and submission script headers. We had to port the scripts to the Torque scheduler, and therefore had to change these details in order to run it on Stallo. There is therefore clearly a need to abstract away from a specific cluster to make it easier to port and configure META-pipe.execution on another cluster (or cloud).

Second, functional analysis is also designed to manage its own workflow, and to submit its own jobs to Stallo by submitting shell scripts. A better solution would be to separate these concerns by generating an intermediate representation of the pipeline and then use it to generate parallel jobs, or to use a workflow manager like Spark.

Third, it uses asynchronous execution, so the start script ends immediately after submitting the jobs to Stallo. This is different from a Galaxy tool as Galaxy tools are expected to run until a job is complete.

Fourth, the generated tool shell scripts do not contain sufficient logic to determine if a job has failed or succeeded; there are numerous ways for which any of the shell scripts or the tools themselves could fail.

In order to resolve these issues and make the integration reliable we developed a Python script that automates the process of inferring the status of a functional analysis job. The script verifies the existence of output files, `grep` log directories for errors and checks cluster status via DRMAA. The Python script also implements a method to wait for all functional analysis tasks to complete. With Spark this approach would not be necessary as Spark already detects failed tasks automatically and can also wait for jobs to finish.

In addition, we experienced that job execution time was limited by straggler nodes. Our new version of META-pipe addresses this issue by decreasing the size per task and increasing the number of tasks, so we can schedule fewer tasks on these straggler nodes.

### 4.3 Galaxy as a user interface

We use Galaxy as the user interface for NeLS users. However, Galaxy has two main issues for a service used by many users. First, Galaxy is a complex system (with more than 300,000 lines of code). This makes it harder to maintain the systems, and harder to understand failures. The latter is especially hard due to limitations in the log output of Galaxy.

The Galaxy Project provides LWR/Pulsar for execution of tools on a remote a machine, which makes the integration with HPC job management systems easier. However, we found it time consuming to develop and especially debug parallel implementations of META-pipe tools. Remote execution also makes data reuse between tools difficult. Since we implemented the functional analysis as a single Galaxy tool, we can reuse the intermediate data between pipeline tool runs. If we had a more pure Galaxy implementation, with all functional analysis stages implemented as Galaxy tools, we would have a larger overhead due to the need to copy data from Galaxy and the cluster before and after the execution of each tool.

The alternative to a flexible workbench interface as provided by Galaxy and Taverna (50) is a custom web application interface. The user cannot easily modify the pipeline, nor integrate it with other analyses. But, it allows tailoring the interface for META-pipe, and the reduced software complexity makes it easier to predict resource usage and handle failures. We will therefore use a web application as the user interface for our publically available META-pipe service that we are currently developing.

### 4.4 Alternative Infrastructures

We have initially focused our efforts into integrating META-pipe with Norwegian infrastructure resources. We plan to integrate META-pipe with compute (such as EGI Federated Cloud (51), EMBL Embassy Cloud (52)) and storage resources (such as EUDAT) provided in the ELIXIR infrastructure, and use ELIXIR services such as AAI, data transfers, and tool registries.

In addition to research infrastructures, commercial cloud computing platforms are an emerging platform for biological data processing. Commercial clouds store large-scale biological datasets (as in Amazon

AWS (53)), and provide the compute resources for analyzing the datasets (e.g. Amazon EC2). Cloud services such as EC2 run virtual machines provided by the user in very large data-centers. The user only pays for the resources needed. There are three main advantages of cloud computing for large-scale biological data. First, the user gets access to a very large compute cluster designed for data-intensive computing. Second, the cloud provides the resources and elasticity needed to scale the job for very large datasets. Third, the processing can be done close to data. A disadvantage is the cost of resources, which is often higher than in research infrastructure centers. In addition, cloud platforms are generally not as optimized for scientific computing programs, with low-latency and high-bandwidth communication requirements, as HPC platforms. But our results show that META-pipe does not benefit from HPC platforms.

Galaxy provides the CloudMan (54) extensions to execute jobs on cloud infrastructure, such as Amazon EC2. There are also many systems that can be used to run biological data processing pipelines such as SLURM (55), Open Lava (56) and Condor (57). MapReduce (58) and similar application frameworks such as Spark (59), are an alternative both for distributing tools for execution, as well as distributed processing. These systems provide a different programming model, scale well to very large datasets, and they handle load balancing and fault tolerance. However, these are typically not integrated with Galaxy, and hence require an integration approach as described in this paper.

Illumnia offers the infrastructure Basespace (60), which is a cloud-based platform for next-generation sequencing data management and analysis. Users can store and share sequencing data, and simplify and accelerate data analysis via the integrated web-based interface. Independent labs can also set-up and monitor their sequencing runs in real time on their Illumina instruments. However, it does not offer extensive analyses and annotation of full-length genes and using Basespace is costly.

We have also built infrastructure systems to reduce the resource usage of META-pipe in computer science research projects, but these are currently not used by the production version of META-pipe. GeStore (61) is a system for adding transparent incremental updates to biological data processing pipelines. We use GeStore to periodically update large-scale compendia and to manage large-scale reference databases such as UniProt. We built GeStore since the processing time for a full compendium update can be several days even on a large computer cluster, making it impractical to frequently update large-scale compendia. We have achieved a 14-fold speedup in analysis time for dataset updates when using GeStore with META-pipe. Mario (62) is a system designed to interactively tuning pipeline tool parameters using fine grained iterative processing of the META-pipe data.

## 5 Conclusions

The Galaxy based META-pipe is a powerful analysis pipeline for metagenomic samples which is intuitive and easy to use for biologists without extensive programming competence. META-pipe is flexible, modular, and it is integrated with large-scale computer systems and identity providers needed to operate a service with a large user base.

META-pipe is currently freely available for all Norwegian academic users, and by requests for other users. These users can also apply for storage quotas on Norwegian research infrastructure. We are currently integrating META-pipe with ELIXIR (63) and other European infrastructure resources. In addition, we are implementing a new version of META-pipe that has a Spark backend, a web application user interface, and a layered architecture that makes it easy to port to new execution environments. The new version will be publically available and tightly integrated with infrastructure systems used by European academic users.

META-pipe is open sourced at: https://github.com/emrobe/META-pipe

# 6 Acknowledgments